\documentclass{article}

\usepackage{arxiv}

\usepackage[utf8]{inputenc} 
\usepackage[T1]{fontenc}    
\usepackage{hyperref}       
\usepackage{url}            
\usepackage{booktabs}       
\usepackage{amsfonts}       
\usepackage{nicefrac}       
\usepackage{microtype}      
\usepackage{lipsum}
\usepackage{graphicx}
\usepackage{amsmath}
\usepackage{multirow}
\usepackage{algorithm}
\usepackage{algorithmic}
\graphicspath{ {./images/} }

\title{Alleviating User-Sensitive bias with Fair Generative Sequential Recommendation Model}

\author{
 Yang Liu \\
  School of Information Management\\
  Nanjing University\\
  Nanjing, 210023 \\
   \And
 Feng Wu \\
  Saw Swee Hock School of Public Health\\
  National Univeristy of Singapore\\
  Singapore, 117549 \\
  \And
 Xuefang Zhu\footnote{Corresponding author} \\
  School of Information Management\\
  Nanjing University\\
  Nanjing, 210023 \\
  \texttt{xfzhu@nju.edu.cn } \\
}

\begin{document}
\maketitle
\begin{abstract}
Recommendation fairness has recently attracted much attention. In the real world, recommendation systems are driven by user behavior, and since users with the same sensitive feature (e.g., gender and age) tend to have the same patterns, recommendation models can easily capture the strong correlation preference of sensitive features and thus cause recommendation unfairness. Diffusion model (DM) as a new generative model paradigm has achieved great success in recommendation systems. DM's ability to model uncertainty and represent diversity, and its modeling mechanism has a high degree of adaptability with the real-world recommendation process with bias. Therefore, we use DM to effectively model the fairness of recommendation and enhance the diversity. This paper proposes a \textbf{Fair} \textbf{GEN}erative sequential \textbf{Rec}ommendation model based on DM, FairGENRec. In the training phase, we inject random noise into the original distribution under the guidance of the sensitive feature recognition model, and a sequential denoise model is designed for the reverse reconstruction of items. Simultaneously, recommendation fairness modeling is completed by injecting multi-interests representational information that eliminates the bias of sensitive user features into the generated results. In the inference phase, the model obtains the noise in the form of noise addition by using the history interactions which is followed by reverse iteration to reconstruct the target item representation. Finally, our extensive experiments on three datasets demonstrate the dual enhancement effect of FairGENRec on accuracy and fairness, while the statistical analysis of the cases visualizes the degree of improvement on the fairness of the recommendation.
\end{abstract}



\maketitle

\section{Introduction}
Sequential recommendation aims to exploit users' historical interactions to mine potential interests and then predict possible future interaction items. With the development of deep learning, Recurrent Neural Networks \cite{hidasi2018recurrent, hidasi2016parallel, quadrana2017personalizing}, Transformer \cite{kang2018self, sun2019bert4rec}, Convolutional Neural Networks \cite{tang2018personalized, yuan2019simple}, and Graph Neural Networks \cite{wu2019session, ma2020memory} have been successfully applied to sequential recommendation with significant results. However, optimizing recommendation accuracy alone may lead to unfairness; one of the important elements is the bias in recommendation pattern due to user-sensitive features (e.g., gender and age) that are prone to discrimination. For example, in the movie recommendation system, men are often recommended thriller movies, while women tend to receive recommendations for romance films. With the accumulation of user interactions and participation in model training, user interests with gender tendencies will be easily captured by the model, and the recommendation bias caused by user-sensitive features will be generated.
Therefore, this recommendation mode seriously affects the experience of users with diverse interests, and the unfairness caused by this phenomenon limits the diversity of recommendation results.

User's potential interests are closely related to recommendation fairness, i.e., users with broader and more segmented interests are able to see the content that meets their needs without being influenced by the group; then, it is the embodiment of fairness from the user's perspective. However, in real recommendation scenarios, faced with the diversity and dynamics of user interests \cite{fan2021modeling, zhou2010solving}, traditional sequential recommendation methods usually characterize the items as fixed vectors. Their singularity and staticity limit the ability to express diversity and uncertainty.

Diffusion model (DM) \cite{ho2020denoising}, as a generative model, was first proposed in image generation. Due to sequential recommendation is essentially based on the biased historical interactions to infer the future interaction probability of items, the noise reduction mechanism of DM is highly adaptable to the task. Additionally, some works of DM in generative recommendation have verified its ability in the mining of user's multi-interests and the diversity of item representation \cite{li2023diffurec}, which assists in getting rid of the "group-specific impression" that can be easily captured by the model. DM's advantages in uncertainty modeling can help users break through the constraints of the fixed pattern of the group. 

Based on the above, DM is very suitable for modeling sequential recommendation and solving recommendation unfairness. Therefore, this paper proposes a \textbf{Fair} \textbf{GEN}erative sequential \textbf{Rec}ommen dation model, FairGENRec, which is built for recommendation fairness by combining DM with sequential recommendation. In the training phase, we propose a sensitive feature recognize model based on warm-up schema training, which is used to guide FairGENRec to incrementally inject random noise into user interactions to obtain corrupted representation, avoiding adding noise to obtain pure Gaussian noise that is completely divorced from the information of personalized user-sensitive features. The sequential denoise model provides a mapping from the noisy state to the original state in the reverse process. In the inference phase, to retain the personalized information, we mimic the forward noise addition process to noisily characterize the historical interaction sequences and then be used for generating items. 

Moreover, from the vantage point of recommendation equity, we have developed a sequential denoise model based on the Transformer architecture. This model incorporates a routing mechanism designed to excavate the multifaceted interests manifest in user interactions. It is seamlessly integrated with user-side modeling to facilitate a two-stage training process. In the initial phase, this approach systematically mitigates biases associated with user-sensitive features within these interests. Subsequently, the refined multi-interests representations are employed to adjust the outputs of the sequential denoise model, thereby aligning the modeling process more closely with the principles of recommendation fairness. The second phase leverages user-insensitive features to cultivate a personalized model, thereby augmenting the personalization of recommendation. Our contributions can be summarized as follows:
\begin{itemize}
    
    \item We proposed a fair generative sequential recommendation model. By combining DM and sequential recommendation, we design the forward diffusion process that guides the real item adding noise through sensitive feature recognize model;
    \item We propose a fairness modeling process that combines the multi-interests module with two-stage training. It can help eliminate bias towards sensitive user features during the denoising process. To the best of our knowledge, this is the first work that attempts to introduce DM into recommendation fairness;
    \item We conduct extensive experiments on three real-world datasets to demonstrate the significant improvement effect of FairGENRec over baselines in both accuracy and fairness.
\end{itemize}
\section{Related Work}

\subsection{Sequential Recommendation}
Due to the advantage of Markov chains in sequence dependencies, early stage sequential recommendation work modeled historical interaction sequence as Markov decision processes \cite{shani2005mdp, rendle2010factorizing, he2017translation}; with the development of deep learning, recurrent neural networks and their variations \cite{chung2014empirical, hochreiter1997long} were used to deal with sequential recommendation \cite{hidasi2018recurrent,hidasi2016parallel,quadrana2017personalizing}; after that, some work analogized them to images for matrix characterization, proving the effectiveness of CNN on sequential recommendation \cite{tang2018personalized,yuan2019simple}, meanwhile, some works used GNN to model and capture complex transformations of historical interaction sequences \cite{wu2019session,ma2020memory}; due to the emergence of self-attention \cite{vaswani2017attention}, Transformer became the mainstream modeling approach for sequential recommendation \cite{kang2018self,sun2019bert4rec}. 
In this paper, we propose a generative sequential recommendation model, which is more relevant to the uncertain behavior of users in real scenarios and promotes recommendation accuracy and diversity.

\subsection{Generative recommendation}
In the recommender system, generative models \cite{kingma2013auto, goodfellow2020generative, ho2020denoising} enhance the diversity of recommendation by modeling users' uncertain potential preferences. Recently, VAE \cite{khattar2019mvae, xie2021adversarial} and GAN \cite{bharadhwaj2018recgan, ren2020sequential} have been widely used in the recommender systems, but both of them have the problems of posterior collapse \cite{lucas2019understanding,zhao2019infovae} and model collapse \cite{salimans2016improved, kingma2016improved}, which lead to poor quality of the generated recommendation. Diffusion model (DM) has better adaptability in generative recommendation due to its strong expressive ability, e.g., DiffRec \cite{wang2023diffusion} combines DM and VAE to achieve personalized recommendation with significant effectiveness; DiffuRec \cite{li2023diffurec} uses DM to model potential aspects of items and user intentions as distributions, and DM has been used in rerank \cite{lin2023discrete} and data augmentation \cite{liu2023diffusion}. In this paper, our work applies DM to sequential recommendation to achieve a fairer generative sequential recommendation model.

\subsection{Fairness-aware Recommendation}
Since recommender systems are highly data-driven,it is essentially a self-reinforcing feedback loop, i.e., if biased data is used for model training, the model may generate unfair recommendations \cite{li2023fairness}. 
Most works promote recommendation fairness by regularizing constraints \cite{li2021user, yao2017beyond}; another approach learns fair representation through adversarial learning via a min-max game between the main task predictor and the adversarial classifier \cite{beigi2020privacy, wu2021fairness}. In addition, some studies model the recommendation problem as a markov decision process and solving the recommendation unfairness through reinforcement learning \cite{ge2022toward, liu2021balancing}. Recently, causal methods have been used to study the causal relationship between sensitive variables and recommendation \cite{wu2018discrimination,zheng2021disentangling}. In this paper, we achieve recommendation fairness by incorporating interest representation that eliminates user-sensitive bias in generative recommendation mechanisms.
\vspace{-2mm}
\begin{figure*}[tb]  
	\centering
	\includegraphics[width=1\textwidth]{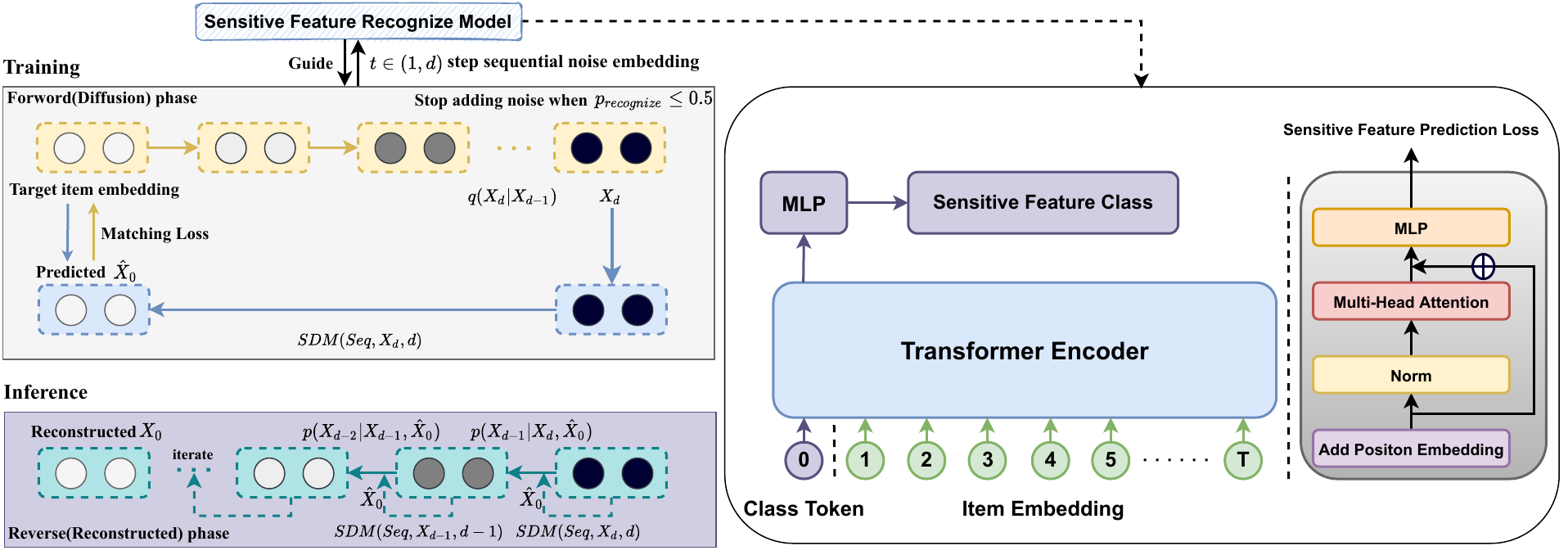}
	\caption{The figure illustrates the forward (diffusion) phase for training and the reverse (reconstructed) phase for inference.}
 \vspace{-2mm}
	\label{fig:traning_inference}
\end{figure*}
\section{Preliminary}\label{pre}
In this section, we will give a brief introduction to DM \cite{ho2020denoising}, including the forward process, the reverse process and optimization process.

\textbf{Forward Process}. Assuming we have an original input  $x_0 \in \mathcal{R}^d$. In the forward process, DM completes the diffusion by gradually adding $T$ times Gaussian noise to $x_0$, forming a Markov chain $x_{1:T}$. Specifically, each transition of $x_{t-1}\rightarrow x_t$ follows: $q(x_t \mid x_{t-1})=\mathcal N(x_t;\sqrt{1-\beta_t}x_{t-1},\beta_tI)$, where $t$ refers to the diffusion step, and $\beta_t \in (0,1)$ is the noise scale at the t-th diffusion step.

\textbf{Reverse Process}. DM denoises completely perturbed $x_T$ in an iterative way to recover the original distribution 
$x_0$. Specifically, each $x_{t-1}$ can be obtained based on $x_0$ and $x_t$ as follows: 

\begin{equation}
\label{4-4}
    q(x_{t-1} \mid x_t,x_0)=\mathcal{N}(x_{t-1};\mu_t(x_t,x_0),\frac{1-\overline{\alpha}_{t-1}}{1-\overline{\alpha}_t}\beta_tI),
\end{equation}
where $\mu_t(x_t,x_0)=\frac{\sqrt{\overline{\alpha}_{t-1}}\beta_t}{1-\overline{\alpha}_t}x_0+\frac{\sqrt{\overline{\alpha}_t}(1-\overline{\alpha}_{t-1})}{1-\overline{\alpha}_t}x_t$, $\overline{\alpha}_t=\prod_{i=1}^t\alpha_i $ and $\alpha_i=1-\beta_i$. Since $x_0$ is unknown, DM estimates 
$\hat{x}_0 \approx x_0$ with the help of a neural network, such as U-Net \cite{ronneberger2015u}.

\textbf{Optimization Process}. DM optimizes parameters by minimizing the Variational Lower Bound (VLB) \cite{sohl2015deep}:  
\begin{equation}
\begin{aligned}
        \mathcal{L}_{vlb}=&\underbrace{E_q[D_{KL}(q(x_t\mid x_0)||p(x_t))]}_{L_T}-\underbrace{logp_\theta(x_0 \mid x_1)}_{L_0}\\&+\underbrace{E_q[\sum_{i=2}^tD_{KL}(q(x_{i-1}\mid x_i,x_0)||p(x_{i-1}|x_i))]}_{L_{t}},
\end{aligned}
\end{equation}
where $p(x_t)$ obeys the standard Gaussian distribution $\mathcal{N}(0,1)$ and the objective function is divided into three parts: 1) $L_T$ in order to make the $q(x_t\mid x_0)$ obtained by forward noise addition approximate the standard Gaussian distribution; 2) $L_0$ optimizes the negative log-likelihood of the final prediction; 3) $L_{t}$ optimizes the reverse process by minimizing the KL divergence between $q(x_{t-1} \mid x_t,x_0)$ and $p_\theta(x_{t-1}\mid x_t)$ (modeled with neural networks).

For simplification, we can simplify it to the following formula \cite{ho2020denoising}:
\begin{equation}\label{8}
    \mathcal{L}_{simple}=E_{t,x_0,\epsilon}[||\epsilon-\epsilon_{\theta}({\overline{\alpha}_tx_0}+\sqrt{1-\overline{\alpha}_t}\epsilon,t)||^2],
\end{equation}
where $\epsilon$ obeys a Gaussian distribution $\mathcal{N}~(0,I)$ and $\epsilon_{\theta}$ is the predicted noise, and the key of DM is to close the gap between the forward-added noise $\epsilon$ and the predicted noise $\epsilon$. 

\section{Method}
\subsection{Problem Formulation}
In this paper, we define $\mathcal{U}$ as the user set and $\mathcal{I}$ as the item set. For each user $u\in \mathcal{U}$, we organize the interactions into a sequence $\mathcal{S}=[i_1^u,i_2^u,\cdots,i_T^u]$ by sorting them in chronological order, where $i_t^u\in \mathcal{I}$ represents the item interacted with by user $u$ at the t-th moment, and $T$ represents the maximum length of the user's historical interaction sequence. Then, the objective of the sequential recommendation model is to recommend the next most likely interaction item $i_{T+1}^u$ for user $u$ based on the historical interaction sequence $\mathcal{S}$.

\subsection{Apply DM to Sequential Recommendation}\label{d2s}
In this paper, we fuse the DM with sequential recommendation, applying forward noise injection and reverse denoise recovery for the target items in the sequence, as shown in Figure \ref{fig:traning_inference}.

\subsubsection{\textbf{Forword (Diffusion) Process.}}\label{421} We first embed the interaction sequence $\mathcal{S}$ into $E$, i.e., $E=Embedding(\mathcal{S})$, where $Embedding \in \mathcal{R}^{|I|*d}$ is the embedding table with hidden dimensionality $d$.

In the diffusion process, we perform forward diffusion for each item embedding $e_t$ in the sequence except for the first item. Specifically, to prevent the original representation from being completely decoupled from the personalized sensitive feature information after being perturbed by random noise, we obtain the diffusion step with the help of the Sensitive Feature Recognize Model (SFRM), which identifies the user-sensitive feature class based on historical interactions. Briefly, the SFRM introduces a learnable class token at the initial position of the sequence and utilizes the Transformer Encoder for global attention computation, and then extracts the output corresponding to the position of the class token for subsequent classification. By feeding the results of each diffusion into the SFRM, the diffusion is stopped when the recognition probability $p_{recognize}\leq0.5$ for the first time, at which time the diffusion step is $d$, note that the diffusion step has the constraint of a maximum value $N$. In addition, in the early stages of training, when SFDR is not well-trained, it may misdirect the diffusion process, so we designed the warm-up schema, i.e., we obtain the diffusion step by random sampling in the early training, and back to the guided mode after the SFRM pretraining is complete, details in section \ref{traininfer}.

To better define the modeling process, we denote $E_n$ as $[e_{2, n}^u, e_{3, n}^u, $ $\cdots, e_{T, n}^u]$, e.g., $e_{t,0}^u$ denotes the initial embedding of the item that user interacts with at the $t\in[2,T]$ moment when diffusion has not occurred. For the forward diffusion equation, see "Forward Process" in section \ref{pre}. Finally, each item $i_t$ is embeded as $e_{t,d}$, representing the noise-added embedding of the item interacted by the user at the $t\in[2,T]$ moment after a diffusion process of $d$ times.

\begin{figure*}[!htb]  
	\centering  
	\includegraphics[width=0.93\textwidth]{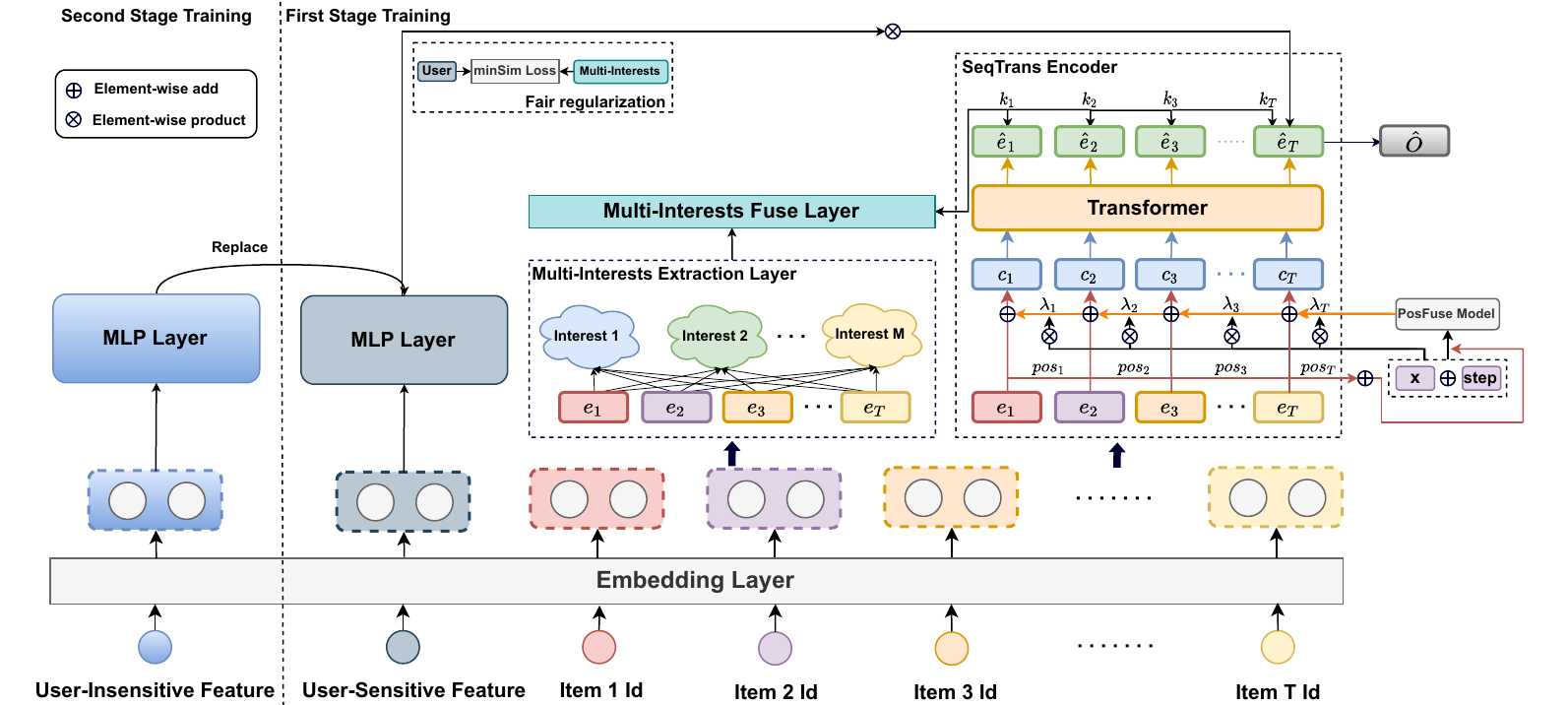}  
	\caption{Overall of proposed sequential denoise architecture.}
	\label{fig2:sdm}  
\end{figure*}

\subsubsection{\textbf{Reverse (Reconstructed) Process.}}\label{422} In the reverse process, we start with the forward noised sequence embedding $E_d$, denoted here as $X_d$, and restore the initial item embedding sequence $X_0$ by iterative denoising. Specifically, we utilize the current reverse step $d$ and $X_d$ as inputs, and obtain the estimated initial embedding $\hat{X}_0$ for the current step through the Sequential Denoise Model (SDM) introduced in section \ref{sdm}. After that, we input $\hat{X}_0$ together with $X_f$ into Eq.\eqref{4-4} to get $X_{d-1}$, and so on, repeatedly iterating in the direction of the step to 0. Finally, $X_0$ after the reverse reduction of DM is obtained, as shown in Algorithm \ref{reverse}.
\begin{algorithm} 
	\caption{\textbf{Reverse (Reconstructed) Process}} 
	\label{reverse} 
	\begin{algorithmic}
		\REQUIRE Initial Embedding $E_0$, Noised Embedding $X_d$, User Feature $f_{u}$, Reverse Step $d$, Sequential Denoise Model $SDM(\cdot)$.
		\ENSURE predicted target item $X_0$. 
		\FOR{$t \leftarrow d,0$} 
		\STATE $\hat{X}_0 = SDM(E_0,X_d,f_{user},f)$; 
		\STATE $X_{t-1}$ is generated by Eq.\eqref{4-4};
		\ENDFOR 
        \RETURN $X_0$.
	\end{algorithmic} 
\end{algorithm}
\subsection{Sequential Denoise Model}\label{sdm}
We need to complete DM training and inference with the help of an auxiliary model. Inspired by the Transformer-based sequential recommendation model \cite{kang2018self,sun2019bert4rec}, we use Transformer \cite{vaswani2017attention} as the basis of the main module, and we perform fairness modeling based on multi-interests to increase the recommendation fairness by eliminating user-sensitive feature bias. The architecture is shown in Figure \ref{fig2:sdm}.
\subsubsection{\textbf{SeqTrans Encoder.}} We use the Transformer as the basis to design the main module for denoising, called SeqTrans Encoder, denoted $f_{\theta}(\cdot)$. The modeling mechanism is as follows:
\begin{equation}\label{11} [\hat{e}_2^u,\hat{e}_3^u,\cdots,\hat{e}_T^u]=f_\theta([c_1^u,c_2^u,\cdots,c_{T-1}^u]),
\end{equation}
\begin{equation}\label{12}
    c_t^u=e_{t,0}^u+\lambda_t\otimes (e_{t+1,d}^u+emb_{step}),t\in[1,T-1],
\end{equation}
\begin{equation}\label{13}
    \lambda_t=gate(e_{t+1,d}^u+emb_{s}+emb_{pos}).
\end{equation}

Suppose the current reverse step is $d$, See Eq.\eqref{11}, where the input of $f_\theta(\cdot)$ is $C=[c_1^u,c_2^u,\cdots,c_{T-1}^u]$, i.e., the initial embedding $E_0$ after the influence of noise embedding $E_d$, and the output is the sequence prediction $\hat{E}=[\hat{e}_2^u,\hat{e}_3^u,\cdots,\hat{e}_T^u]$ for the next moment, which corresponds to the $\hat{X}_0$ mentioned in section \ref{422}. For obtaining $c_t^u$ is shown in Eq.\eqref{12}, taking the item at $t$ moment in the sequence as an example, $e_{t+1,d}^u$ denotes the embedding of the target item after forward diffusion noise-added at $t+1$ moment; $emb_{s}$ denotes the embedding of the reverse step $d$; and $\lambda_t$ controls the extent to which the item embedding at $t$ moment $e_{t,0}^u$ is affected by the noise-added target item, in order to incorporate long and short-term information, and reveal the implied relationship between the moment of occurrence of the item interaction and the target item's recover, we design a PosFuse network to learn $\lambda_t$, which is a two-layer neural network determined by $e_{t+1,d}^u$, $emb_{s}$, and position embedding $emb_{pos}$, with the final layer's activation function being $2*sigmoid$.

\subsubsection{\textbf{Fair Modeling with Multi-Interests.}} We design a multi-interests extraction and fuse module that removes user-sensitive feature bias to correct the results of  \textbf{SeqTrans Encoder}, aiming to promote the recommendation fairness.

\textbf{Multi-Interests Extraction Layer.} Dynamic routing algorithms were initially proposed to address the shortcomings of CNN \cite{sabour2017dynamic}, and have since been applied to the multi-interests recommendation \cite{li2019multi, cen2020controllable}. Specifically, the multi-interests extraction layer takes the user sequence embedding $E$ as input, and the output a set of capsule interest vectors $V_u=[v_1^u,v_2^u,\cdots,v_M^u]$ that can represent the user's diverse intents through the dynamic routing algorithm, see details in \cite{li2019multi}.

\textbf{Multi-Interests Fuse Layer.} After extracting multi-interests, we use the target item embedding and the capsule interest vectors to complete the Label-aware Attention, see Eq.\eqref{14}, where $p$ is the index of pow, and finally get the multi-interests embedding $k_{target}^u$:
\begin{equation}\label{14}
    k_{target}^u=V_usoftmax(pow(V_u^Tk_{target}^u,p)).
\end{equation}

We use the output of the SeqTrans Encoder $\hat{E}$ as the target item embedding for the operation of Label-aware Attention. The final $k_{target}$ obtained represents the user's multi-interests embedding of the results at the moment of predicted by the SeqTrans Encoder for the item at moment $t\in[1,T-1]$ in the sequence.

\textbf{Fair regularization}. To eliminate user-sensitive feature bias in multi-interests, we design the personalized user tower and auxiliary loss to correct the multi-interests embedding $k_{target}$. Model training is divided into two stages, see section \ref{traininfer}, and the bias elimination task is accomplished in the first stage. The features from personalized user tower in this stage are adopted as user-sensitive features, and the design of the sensitive features for the datasets is shown in Table \ref{table:2}. We denote the output of the personalized user tower as $e_{sen}^u$, and then combine multi-interests embedding $k_{t}^u$ to complete the regularization operation as shown in Eq.\eqref{15}, which is used as an auxiliary loss to complete the bias elimination for multi-interests:
\begin{equation}\label{15}
    \mathcal{L}_I=\frac{1}{|\mathcal{U}|}\sum_{u\in \mathcal{U}}\frac{1}{T-1}\sum_{t=2}^T|\frac{k_t^u\cdot e_{sen}^u}{||k_t^u||\cdot||e_{sen}^u||}|.
\end{equation}

We use the multi-interests weighting method to improve recommendation fairness, and add user information to model personalization. Finally, we get the output $\hat{O}$ by:
\begin{equation}\label{16}
    \hat{O}=\hat{E}\otimes k_t^u \otimes e_{sen}^u,
\end{equation}
in the second stage of training, we replace $e_{sen}^u$ with $e_{insen}^u$, which is insensitive user embedding.

\subsection{Model Training and Inference}\label{traininfer}
\subsubsection{\textbf{Two-stage Training.}} To balance recommendation fairness and personalization, we design a two-stage training approach.

Since the length of some user sequences doesn’t satisfy the maximum sequence length defined by FairGENRec, we padded the sequences $\mathcal{S}^u$ (except for the last item), and based on the item’s sequence position, we defined the following for the desired output: 
\begin{equation}
    o_t^u=
    \begin{cases}
    None& \text{if $i_t^u$ is a padding item,}\\
    e_{t+1,0}^u& \text{$1\leq t \leq T-1$}.
    \end{cases}
    \label{eq:10}
\end{equation}

To enhance training efficiency, we negatively sampled each item in $\mathcal{S}^u$ (except the last one) once, denoted as $G=[g_1^u,g_2^u,\cdots,g_{T-1}^u]$, where $G$ and $O$ do not have duplicate items. In addition, the sequential recommendation often uses the potential embedding of the items for inner-product matching to design the loss function, so we choose the binary cross entropy loss as the main loss function, and we will ignore the value of the loss at $o_t^u=None$:
\begin{small}
    \begin{equation}
    \mathcal{L}_M=-\frac{1}{|\mathcal{U}|}\sum_{u\in \mathcal{U}}\frac{1}{T-1}\sum_{t=1}^{T-1}[log(\sigma(\hat{o}^u_t\cdot o^u_t))+log(1-\sigma(\hat{o}^u_t\cdot g^u_t))].
\end{equation}
\end{small}

In addition, we train the SFRM using cross entropy loss:
\begin{equation}
    \mathcal{L}_F=-\frac{1}{|\mathcal{U}|}\sum_{u \in \mathcal{U}}\sum_{i=1}^C f_i^ulog(\hat{f}_i^u),
\end{equation}
where $F=[f_1^u, f_2^u,\cdots,f_C^u]$ represents the label of the sensitive feature, $\hat{F}=[\hat{f}_1^u, \hat{f}_2^u,\cdots,\hat{f}_C^u]$ is the output of SFRM after softmax operation and $C$ is the number of sensitive feature categories.

The first stage adopts co-training, which is responsible for optimizing the SDM and the SFRM. The loss function is as follows:
\begin{equation}
\mathcal{L}=\mathcal{L}_M+\lambda_I\mathcal{L}_I+\lambda_F\mathcal{L}_F,   
\end{equation}
where $\lambda_I$ and $\lambda_F$ control the importance of their corresponding losses. In the second stage, we completely abandon the personalized user tower based on user-sensitive features, and instead model user personalization using user-insensitive features. Note that the main loss function is used to train the newly added personalized user tower. Combined with the warm-up schema in section \ref{421}, the model training is shown in algorithm \ref{train}.

\begin{algorithm} 
	\caption{\textbf{Training Process}} 
	\label{train} 
	\begin{algorithmic}
		\REQUIRE Initial Embedding $E_0$, Warm-up Epochs $k$, Learning Epochs $e$, Maximum Diffusion Step $N$, User-Sensitive Feature $f_{sen}^u$, User-InSensitive Feature $f_{insen}^u$, Sensitive Feature Recognize Model $SFRM(\cdot)$,  Sequential Denoise Model $SDM(\cdot)$, Train Stage $ts$, Initialize $i=0$.
		\ENSURE Well-trained $SFRM(\cdot)$ and $SDM(\cdot)$.
        \WHILE{$i \leq e$}
        \IF{$i \leq k$ \&\& $ts == 1$}
        \STATE $d$ is sampled from a uniform distribution $U(1,N)$
        \STATE $E_{d}$ is generated by Eq.\eqref{eq:10};
        \ELSE
        \STATE $p_{recognize} \leftarrow SFRM(E_0)$, $d = 0$;
        \WHILE{$p_{recognize} \leq 0.5$ \&\& $d \leq N$}
        \STATE $E_{d+1}$ is generated by Eq.\eqref{eq:10};
        \STATE $p_{recognize} \leftarrow SFRM(E_{d+1})$, $d = d+1$;
        \ENDWHILE 
        \ENDIF
        \IF{$ts == 1$}
        \STATE $\hat{X}_0 \leftarrow SDM(E_0,E_d,f_{sen}^u,f)$;
        \STATE update parameter for all modules: $\mathcal{L}=\mathcal{L}_M+\lambda_I\mathcal{L}_I+\lambda_F\mathcal{L}_F$;
        \ELSE
        \STATE $\hat{X}_0 \leftarrow SDM(E_0,E_d,f_{insen}^u,f)$;
        \STATE only update parameter for personalized user tower: $\mathcal{L}_M$;
        \ENDIF
        \STATE i = i+1;
        \ENDWHILE
	\end{algorithmic} 
\end{algorithm}

\subsubsection{\textbf{Inference.}} In the inference phase, we use SDM in conjunction with the method in section \ref{422} to obtain the final embedding, denoted as $R$, and then use the last item $r_T$ to perform inner product matching with item embedding $e_i$ in candidate set:
\begin{equation}
    \mathop{\arg\max}\limits_{{i\in I}} Rounding(r_T)=r_T\cdot e_i.
\end{equation}

In the recommendation system, to preserve personalization \cite{zhu2022personalized}, we mimic the forward diffusion process to noisily characterize the average pooling of valid items embedding in the sequence.

\section{Experiments}
We conduct extensive experiments on three real-world datasets with the aim of addressing the following three questions: RQ1: How does our FairGENRec perform compared to the state-of-the-art methods? RQ2: How do the designs of FairGENRec affect the performance? RQ3: From the perspective of case study, how does our FairGENRec perform in recommendation fairness?

\subsection{Experimental Settings}
\subsubsection{\textbf{Datasets.}} We conducted experiments on three publicly available real-world datasets. 1) The Amazon review dataset\footnote{https://jmcauley.ucsd.edu/data/amazon/index\_2014.html} comprises 142.8 million reviews spanning from May 1996 to July 2014, focusing specifically on the beauty category for our analysis; 2) The Movielens dataset contains users’rating on movies, we use ML-1M\footnote{https://files.grouplens.org/datasets/movielens/ml-1m.zip} with about 1 million ratings; 3) Yelp\footnote{https://www.yelp.com/dataset}, a widely used business dataset, contains user reviews covering various restaurants.
\begin{table}[bth]
\centering
\caption{Statistics of preprecessed datasets.}
\label{table:1}

\begin{tabular}{cccccc}
\toprule 
\textbf{Dataset} & \textbf{\# users}  &  \textbf{\# items}  &  \textbf{\# actions}  &  \textbf{Avg.length}  & \textbf{Sparsity} \\ 
\midrule 
\textbf{Beauty} & 22363 & 12101 & 198502 & 8.9 & 99.93\% \\
\textbf{ML-1M} & 6040 & 3416 & 999611 & 165.5 & 95.16\% \\
\textbf{Yelp} & 104072 & 54056 & 1328015 & 12.8 & 99.98\% \\
\bottomrule 
\end{tabular}
\end{table} 

Building upon \cite{kang2018self,sun2019bert4rec}, we arranged the interactions of each user into the initial sequence sorted by timestamps. We filtered out users with fewer than 5 interactions and items with less than 5 interactions. We also used a leave-one-out strategy to partition the training, validation, and test sets.
Finally, considering the dataset characteristics in Table \ref{table:1}, we set the maximum sequence length to 50 for Beauty, 200 for ML-1M, and 50 for Yelp.

\begin{table}[h]
  \centering
  \caption{User-sensitive features and meanings on datasets.}\label{table:2}

\begin{tabular}{cccccc}
\toprule 
\textbf{Dataset} & \textbf{User Bias-Related Features}  &  \textbf{Meaning} \\ 
\midrule 
\textbf{Beauty} & Average Scores & User Tolerance \\
\textbf{ML-1M} & Genders & Gender-Related Preferences \\
\textbf{Yelp} & User Engagement & Information Cocoon\\
\bottomrule 
\end{tabular}
\end{table}

\begin{table*}[h]
    \centering
    \caption{Recommendation performance and fairness results of three datasets, obtained as an average of 5 experiments with different random seeds. Bold numbers are the best results and underlined numbers are the strongest baselines.}
    \label{table:3}
    \resizebox{\textwidth}{!}{
    \begin{tabular}{ccccccccccc}
    \toprule
    \textbf{Dataset} & \textbf{Purpose}	& \textbf{Metric} & \textbf{SASRec}	& \textbf{BERT4Rec} & \textbf{STOSA} & \textbf{ACVAE} & \textbf{DiffRec} &\textbf{DiffuRec} &\textbf{PFRec}& \textbf{FairGENRec}\\
    \midrule
    \multirow{7}*{Beauty} & \multirow{6}*{Performance$\uparrow$} & N@5 &	0.1784&	0.1350 &	0.1684 &	0.1666  &  0.2055	 &\underline{0.2096}&0.1536 
&	\textbf{0.2102} \\ 
    ~ & ~ & N@10 &	0.2057&	0.1656 &	0.2015  &	0.1905  & 0.2281	 &\underline{0.2332}&0.1848 
&	\textbf{0.2347} \\
    ~ & ~ & N@20 &	0.2191 &	0.1913 &	0.2233 &	0.2146  & 0.2491		 &\underline{0.2555}&0.2130 
&	\textbf{0.2579} \\
    ~ & ~ & H@5 &	0.2611&	0.2019 &	0.2464 &	0.2320  & \underline{0.2812}	 &0.2791&0.2319 
&	\textbf{0.2868} \\
    ~ & ~ & H@10 &	0.3422&	0.2865 &	0.3287  &	0.3115  & 0.3545	 &\underline{0.3580}&0.3331 
&	\textbf{0.3625} \\
    ~ & ~ & H@20 &	0.4329 &	0.3863 &	0.4171  &	0.4073  & 0.4455	 &\underline{0.4510} &0.4338&	\textbf{0.4547} \\
    \cmidrule{2-11}
    ~ & Fairness$\downarrow$ & Macro-F1 & 0.2125 	 & 0.2654	 & 0.1978	 &	0.2013	  & 0.2443	& 0.2114 & \underline{0.1156} &	\textbf{0.1059} \\
    \midrule
    \multirow{7}*{ML-1M} & \multirow{6}*{Performance$\uparrow$} & N@5 &	0.3822 &	0.2522 &	0.2854 &	0.3971  & 0.4373	 &\underline{0.4465} &0.1989 
&	\textbf{0.4543}	 \\
    ~ & ~ & N@10 &	0.4287 &	0.3025 &	0.3279 &	0.4342 & \underline{0.4898}	 &0.4854 &0.2460 
&	\textbf{0.4923}\\
    ~ & ~ & N@20 &	0.4573 &	0.3434 &	0.3632 &	0.4606  & \underline{0.5113}	 &0.5097 &0.2911 
&	\textbf{0.5157}	 \\
    ~ & ~ & H@5 &	0.5215 &	0.3745 &	0.4078 &	0.5397  & \underline{0.6101}	 &0.6070 &0.3063 
&	\textbf{0.6149}	 \\
    ~ & ~ & H@10 &	0.6664 &	0.5373 &	0.5449 &	0.6540  & \underline{0.7291}	 &0.7263  &0.4538 
&	\textbf{0.7318}	 \\
    ~ & ~ & H@20 &	0.7971 &	0.7010 &	0.6829 &	0.7555  & 0.8201	 & \underline{0.8221}  &0.6331 &	\textbf{0.8230}	 \\
    \cmidrule{2-11}
    ~ & Fairness$\downarrow$ & Macro-F1 & 0.4915	 & 0.5036	 & 0.4789	 & 0.4745	  &	 0.4696 & 0.4115  & \underline{0.2412} &	\textbf{0.2205}	 \\
    \midrule
    \multirow{7}*{Yelp} & \multirow{6}*{Performance$\uparrow$} & N@5 &	0.3982 &	0.3392 &	0.4080 &		0.4468  & 0.4647	 &\underline{0.4655}  &0.3718 &	\textbf{0.4739}	\\
    ~ & ~ & N@10 &	0.4510 &	0.3884 &	0.4604 &	0.4904  & \underline{0.5081} &0.5073  &0.4045 
&	\textbf{0.5090} \\
    ~ & ~ & N@20 &	0.4808 &	0.4198 &	0.4912 &	0.5173  & 0.5115	 &\underline{0.5266}  &0.4252 &	\textbf{0.5325} \\
    ~ & ~ & H@5 &	0.5615 &	0.4823 &	0.5693 &	0.6055  & 0.6187	 &\underline{0.6351}  &0.5112 
&	\textbf{0.6372} \\
    ~ & ~ & H@10 &	0.7235 &	0.6397 &	0.7312 &	0.7399  & 0.7512	 &\underline{0.7681}  &0.6116 
&	\textbf{0.7799} \\
    ~ & ~ & H@20 &	0.8302 &	0.7704 &	0.8549 &	0.8444  & \underline{0.8751}	 &0.8718  &0.6925 &	\textbf{0.8780} \\
    \cmidrule{2-11}
    ~ & Fairness$\downarrow$ & Macro-F1 & 0.7212	 & 0.7456	 & 0.7021	 & 0.7101  & 0.7010 & 0.6878	 & \underline{0.4512} & \textbf{0.4249} \\
    \bottomrule
    \end{tabular}}
\end{table*}
Then, we design user-sensitive features that lead to unfair recommendation, see Table \ref{table:2}: 1) For Beauty, the user's tolerance affects the item score level of the recommendation, i.e., the stricter the user is, the more the user will probably be not recommended the item with a low average score which leads to the unfair recommendation. We finally choose the average score of the user to characterize user’s tolerance level; 2) For ML-1M, the gender influences the category tendency of the recommendation, e.g., males are more likely to be learned by the model to like the adventure category, while females tend to prefer the romance category; 3) User engagement represents the frequency of the user's use of the platform, the more loyal the user, the more likely to appear information cocoon phenomenon, which ultimately leads to the recommendation of the category of homogenization, thus, we designed a series of user-sensitive features related to user engagement for Yelp, such as the number of comments and elite user status.

As for the Evaluation Metrics, we choose top-K Hit Radio (HR@K) and top-K Normalized Discounted Cumulative Gain (NDCG@K) as our Evaluation Metrics, where $K \in [5,10,20]$. In addition, in order to improve inference efficiency, we fairly sample 200 negative samples for each candidate sample for the evaluation of the effect of all models. For fairness, following \cite{wu2021fairness}, we use Macro-F1 to measure the fairness, with smaller values representing higher fairness.

\subsubsection{\textbf{Baselines.}}  We compare FairGENRec with seven baseline models, including: 1) \textbf{SASRec} \cite{kang2018self} uses Transformer to mine the correlation between items in behavior sequences; 2) \textbf{Bert4Rec} \cite{sun2019bert4rec} learns a bidirectional Transformer Encoder that fuses the information from both ends of behavior sequences. 3) \textbf{ACVAE} \cite{xie2021adversarial} introduces contrastive learning based on the adversarial variational Bayesian for recommendation; 4) \textbf{STOSA} \cite{fan2022sequential} proposed a stochastic self-attention model for sequential recommendation, which represents the items with Gaussian distributions; 5) \textbf{DiffRec} \cite{wang2023diffusion} combines VAE and DM to accomplish generative recommendation in the form of denoising; 6) \textbf{DiffuRec} \cite{li2023diffurec} uses DM to model potential aspects of items and different user intentions as distributions; 7) \textbf{PFRec} \cite{wu2022selective} relies on attribute-specific prompt-based bias eliminators with adversarial training for fairness recommendation. In addition, we apply the SFRM from this paper to all baseline models and compute Macro-F1 by predicting sensitive features based on the output of baselines for fairness evaluation.

\subsubsection{\textbf{Implementation.}}\label{514} 
The key hyperparameters of the FairGENRec as follows: 1) In terms of combining DM with sequential recommendation, we focus on two hyperparameters: one is the \textbf{maximum diffusion step}, which we have repeatedly experimented in the range of $[1,5,10,15,20,25,30]$ and finally determined to be 15; the other is the \textbf{noise schedule}, which we explored 4 ways: linear \cite{ho2020denoising}, sqrt \cite{li2022diffusion}, mutual information \cite{austin2021structured} and cosine \cite{nichol2021improved}, and finally determined as the mutual information schedule; 2) In the design of SDM, we first set the block number of Transformer Encoder to 4 and adopt the single-head self-attention. Meanwhile, interests number is set to 3; 3) For the general parameters, we set the dropout rate to 0.2, the embedding dimension is set to 128, and all parameters are initialized by the Xaiver normalization distribution.
\subsection{Overall Performance Comparisons (RQ1)}
We compare the performance and fairness of FairGENRec with baselines, as shown in Table \ref{table:3}, and get the following observations: Firstly, compared to Transformer-based recommendation models, generative recommendation models present advantages in both accuracy and fairness metrics due to capabilities in uncertainty modeling and distribution representation learning.

Secondly, in generative recommendation models, since DM has the ability to represent users' multi-interests and adaptability to the real-world recommendation process with bias, DiffRec and DiffuRec achieve optimal results in accuracy metrics. However, compared with the optimal effect of the above two baselines, FairGENRec still has a small degree of improvement in accuracy in addition to a significant improvement in fairness. Specifically, the improvement degree of fairness metric is 49.91\%, 46.42\% and 38.22\%, while the maximum improvement effect of accuracy metrics (NDCG/HR) is 0.94\%/1.99\%, 1.75\%/0.79\% and 1.80\%/1.54\% on Beauty, ML-1M and Yelp, respectively. We attribute fairness superiority in avoiding user-sensitive features that may lead to discrimination in recommendation, then by injecting users' multi-interests into uncertainty modeling, it is possible to use increased personalization to mitigate the performance loss of fairness modeling on accuracy.

Finally, compared with the fairness recommendation model PFR ec, FairGENRec's fairness metric shows an improvement of 8.39\%, 8.58\% and 5.83\% on Beauty, ML-1M and Yelp, respectively, and six accuracy metrics lead across the board. Therefore, we find that FairGENRec pays more attention to the strong correlation between users' real interests and recommended items in fairness modeling, while uncertainty modeling brings strong generalization to unseen data, which achieves both accuracy and fairness improvement.

\subsection{Analysis of FairGENRec (RQ2)}
This section analyzes the influence of the key parameters on the effectiveness of FairGENRec, and further verifies the validity of the model by ablation experiment.

\subsubsection{\textbf{Impact of Hyper-parameter Setting.}} 
\begin{figure}[tbp]  
	\centering  
	\includegraphics[width=\linewidth]{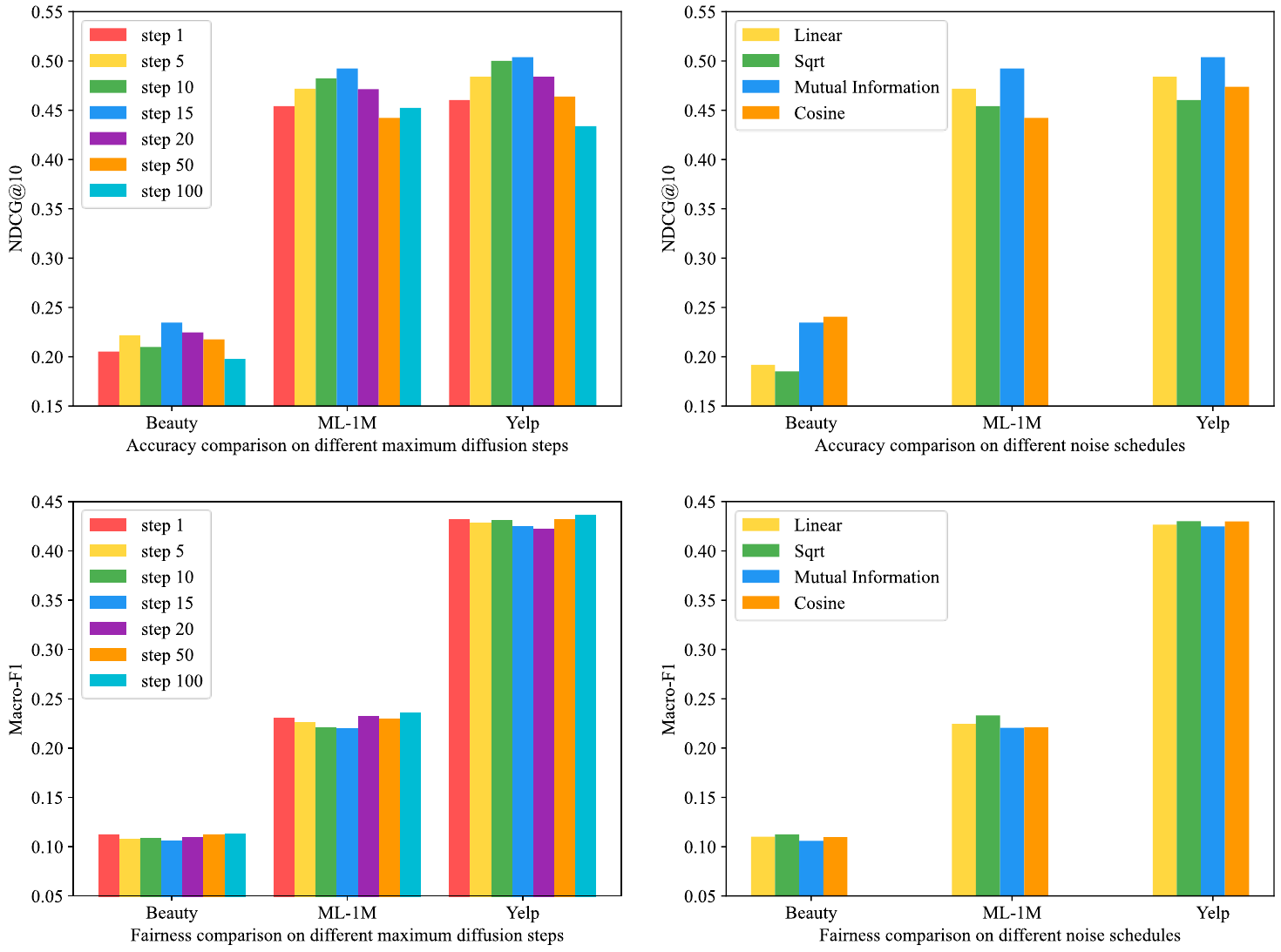}  
	\caption{Performance influence of different maximum diffusion steps and different noise schedules.}
	\label{fig3:beata_step}  
\end{figure}

The \textbf{maximum diffusion step} control the maximum step of forward diffusion, see algorithm \ref{train}, and the two left parts of Figure.\ref{fig3:beata_step} show the performance of FairGENRec on three datasets under different parameter settings, respectively. We find that the maximum diffusion steps should be kept within a reasonable range. If they approach 0, the noise-added effect becomes too weak; conversely, if they increase too much, personalization is lost, inhibiting FairGENRec.

We experimented with four types of \textbf{noise schedules}: 1) \textbf{\textit{Linear Schedule}}: $\beta_t$ increases linearly from $t=1$ to $t=T$ in the range $[10^{-4},0.002]$; 2) \textbf{\textit{Sqrt Schedule}}: This schedule increases noise rapidly at the first few diffusion steps, and then slows down injecting diffusion noises; 3) \textbf{\textit{Mutual Information Schedule}}: to better capture the distributional characteristics of the data, the schedule guides the learning process of the model by linearly interpolating the mutual information between the original data and the hidden variables: $\beta_t=(T-t-1)^{-1}$; 4) \textbf{\textit{Cosine Schedule}}: The schedule utilizes the cosine function to achieve the effect of noise smoothing increase with the formula: $\overline{\alpha}_t=\frac{f(t)}{f(0)},f(t)=cos(\frac{\frac{t}{T}+0.009}{1+0.008}\cdot \frac{\pi}{2})^2$.

The results of noise schedules are shown in the two right panels of Figure \ref{fig3:beata_step}, we find that the mutual information schedule almost achieves the best effect on three datasets, while on Beauty NDCG@10 is slightly worse than the cosine schedule. To pursue global optimization, we choose the mutual information schedule.

    \subsubsection{\textbf{Ablation Study.}} 
\begin{table}[htbp]
    \centering
    \caption{Ablation analysis on three datasets, bold numbers are the best result.}
    \label{table:4}
    \begin{tabular}{ccccccccc}
    \toprule
    \textbf{Dataset}	& \textbf{Type} & \textbf{N@5} &\textbf{ N@10}	& \textbf{N@20} & \textbf{H@5} & \textbf{H@10} & \textbf{H@20} & \textbf{Macro-F1}\\
    \midrule
    \multirow{4}*{\textbf{Beauty}} & \textbf{Default} & \textbf{0.2102} &	\textbf{0.2347} & \textbf{0.2579} &	\textbf{0.2868} &	\textbf{0.3625} &	\textbf{0.4547} &	\textbf{0.1059}\\
    ~ & \textbf{W/O SFRM} & 0.2096 &	0.2332 &	0.2555 &	0.2802 &	0.3535 &	0.4424 &	0.1127\\
    ~ & \textbf{W/O DSE} & 0.2049 &	0.2271 &	0.2483 &	0.2759 &	0.3445 &	0.4281 &	0.1079\\
    ~ & \textbf{W/O UT} & 0.2057 &	0.2267 &	0.2501 &	0.2765 &	0.3417 &	0.4345 &	0.1062\\
    \midrule
    \multirow{4}*{\textbf{ML-1M}} & \textbf{Default} & \textbf{0.4543} &	\textbf{0.4923} & \textbf{0.5157} &	\textbf{0.6149} &	\textbf{0.7318} &	\textbf{0.8230} &	0.2205\\
    ~ & \textbf{W/O SFRM} & 0.4465 &	0.4854 &	0.5097 &	0.6070 &	0.7263 &	0.8219 &	0.2303\\
    ~ & \textbf{W/O DSE} & 0.4299 &	0.4695 &	0.4960 &	0.5950 &	0.7167 &	0.8204 &	0.2231\\
    ~ & \textbf{W/O UT} & 0.4393 &	0.4803 &	0.5049 &	0.5978 &	0.7238 &	0.8200&	\textbf{0.2203} \\
    \midrule
    \multirow{4}*{\textbf{Yelp}} & \textbf{Default} & \textbf{0.4575} &	\textbf{0.5039} & \textbf{0.5271} &	\textbf{0.6351} &	\textbf{0.7775} &	\textbf{0.8739} &	\textbf{0.4249}\\
    ~ & \textbf{W/O SFRM} & 0.4503 &	0.4986 &	0.5266 &	0.6196 &	0.7681 &	0.8718 &	0.4335\\
    ~ & \textbf{W/O DSE} & 0.4411 &	0.4992 &	0.5172 &	0.6187 &	0.7512 &	0.8698 &	0.4278\\
    ~ & \textbf{W/O UT} & 0.4491 &	0.5001 &	0.5189 &	0.6255 &	0.7663 &	0.8701 &	0.4252\\
    \bottomrule
    \end{tabular}
\end{table}
To verify the effectiveness of different components, i.e., SFRM, diffusion step embedding and second stage's personalized user tower, we designed three models: 1) FairGENRec without SFRM (W/O SFRM); 2) FairGENRec without diffusion step embedding (W/O DSE); 3) FairGENRec without second stage training (W/O UT), and compares the performance with the default FairGENRec. Results are shown in Table \ref{table:4}, we find that: 1) SFRM retains the user-sensitive feature personalized information for debias, which mainly promotes recommendation fairness; 2) The diffusion step optimizes the reverse process by providing the model with the current noise level, which has a large impact on recommendation accuracy; 3) The personalized user tower in the second stage provides tailored information to improve recommendation accuracy without significantly affecting fairness.

\subsection{Fairness Analysis (RQ3)}\label{fair}

We statistically analyze recommendation results from FairGENRec and DiffuRec against the sensitive features of each dataset to demonstrate the promotion of FairGENRec on recommendation fairness.

\begin{figure}[htb]  
	\centering  
	\includegraphics[width=0.7\textwidth]{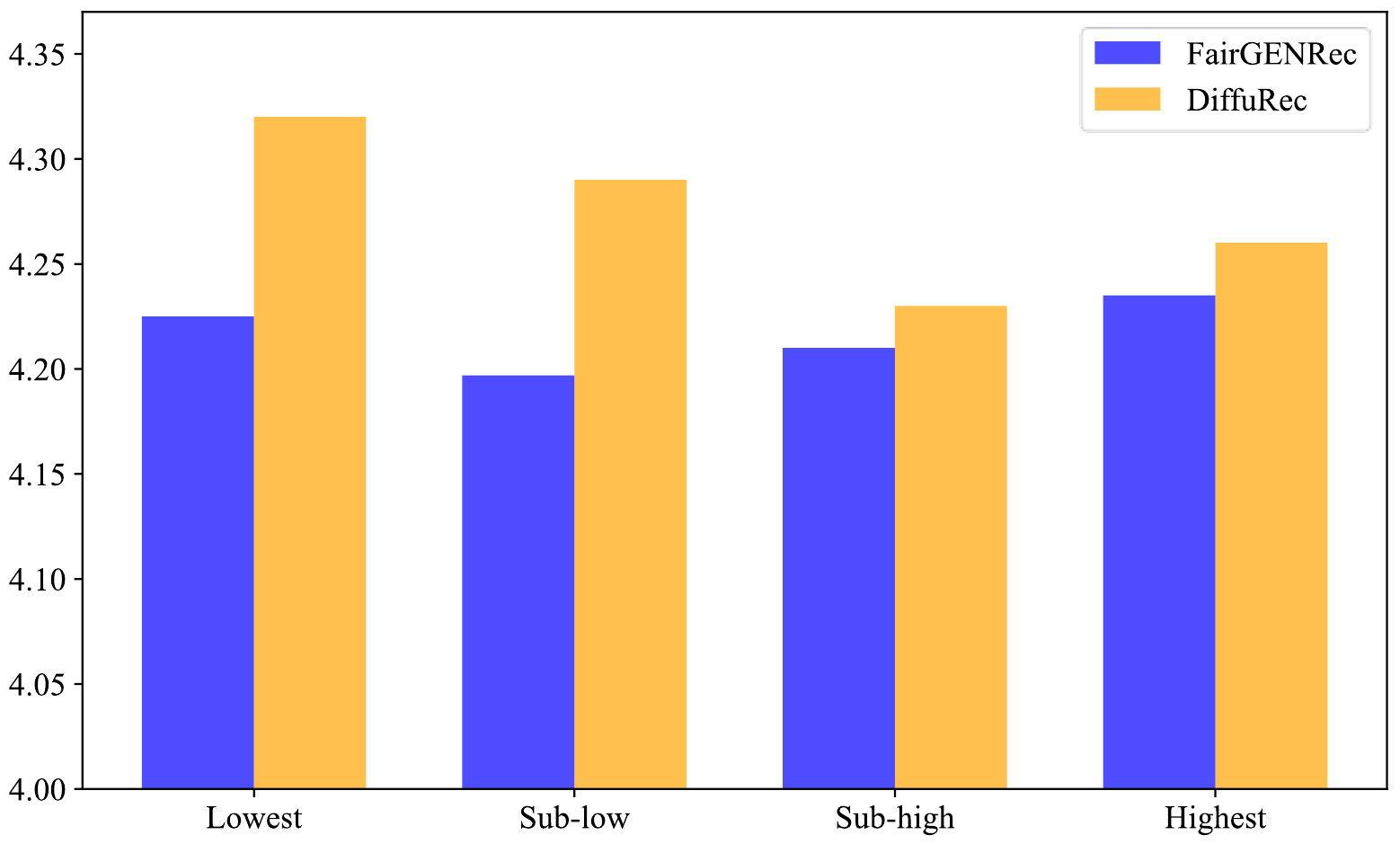}  
	\caption{Fairness comparison on Beauty.}
	\label{fig4:beauty}  
\end{figure}

For Beauty, we divide the user's tolerance level into four levels according to historical average rating, which are Lowest, Sub-low, Sub-high and Highest, and then count the average ratings of recommended items by FairGENRec and DiffuRec under different user's tolerance levels respectively. The results are shown in Figure \ref{fig4:beauty} and we find that: 1) DiffuRec's statistical results show that there is an almost inverse correlation between the user tolerance level and the rating level of the recommendation, i.e., the lower the tolerance level is, the higher the rating level of the items recommended is;
2) The overall rating level of recommendation in FairGENRec decreases compared to DiffuRec, especially for users with lower tolerance level, which reflects that FairGENRec significantly improves the recommendation bias for this type of users. Meanwhile, it is more friendly to the items with low ratings.

\begin{figure}[htb]  
	\centering  
	\includegraphics[width=\linewidth]{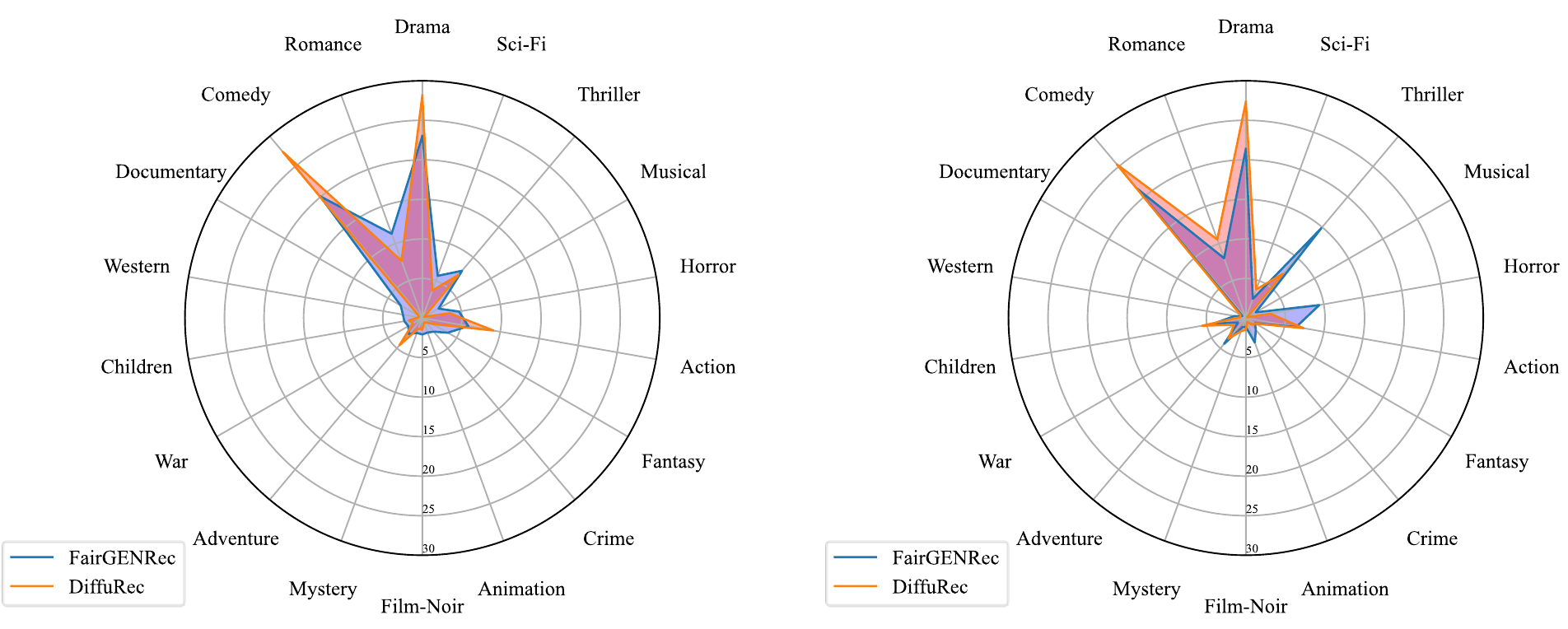}  
	\caption{Fairness comparison on ML-1M.}
	\label{fig5:ml-1m}  
\end{figure}

\begin{figure}[htb]  
	\centering  
	\includegraphics[width=\linewidth]{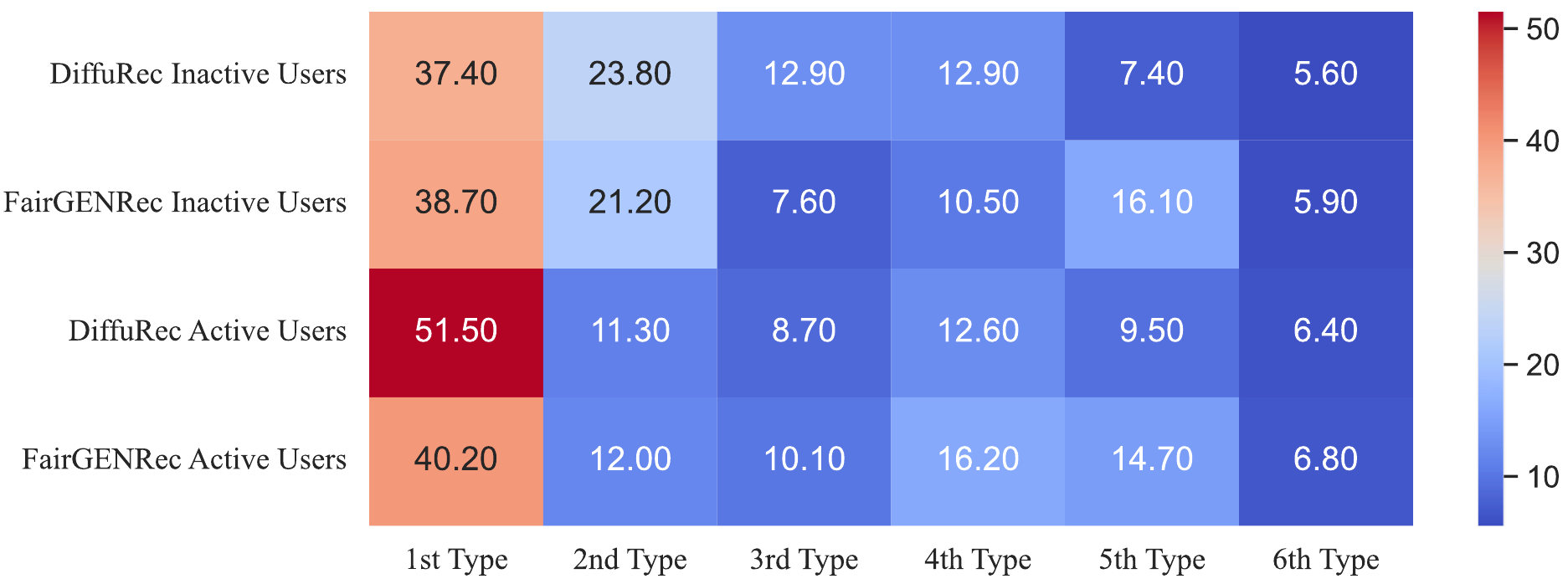}  
	\caption{Fairness comparison on Yelp.}
	\label{fig6:yelp}  
\end{figure}

On ML-1M and Yelp, FairGENRec pays more attention to prevent the recommendation from being homogeneous by eliminating bias. For ML-1M, We use gender as the classification basis to count the recommended item categories for ML-1M. For Yelp, we use the number of user interactions to differentiate the degree of user engagement and use it as the basis for the classification, and we find that: 1) The left and the right graphs of Figure \ref{fig5:ml-1m} respectively show the category statistics of recommended items for male and female on ML-1M by FairGENRec and DiffuRec. Although Comedy and Drama still occupy the mainstrea categories, FairGENRec shows a decrease compared to DiffuRec. Meanwhile, FairGENRec recommends fewer movies for males in genres such as action and adventure movies, while increasing its recommendation of romance and children, which are more likely to be liked by females. In addition, it increases the number of movies it recommends to females, such as thriller and horor. 
2) Figure \ref{fig6:yelp} reflects the category statistics of FairGENRec and DiffuRec's recommendation for both active and inactive users, and the seven categories are obtained by text clustering according to “business categories”. Specifically, there is a difference from the same model for active and inactive users, e.g., the recommendation ratio of the 1st type of DiffuRec for active users and inactive users is 51.50\% and 37.40\% respectively, which proves that the information cocoon phenomenon of active users is more frequent. Compared to DiffuRec,  FairGENRec also has the mitigating effect on the bias, e.g., FairGENRec reduces the recommendation ratio of 1st type for active users by 11.3\%, and the recommendation ratio for seven types is more average, but the optimization of recommendation ratio for inactive users is not obvious.

\section{Conclusion}
In this work, considering recommendation unfairness caused by user-sensitive features, we propose a generative sequential recommendation model called FairGENRec, which is based on the DM and achieves generative recommendation of items by injecting random noise under the guidance of the sensitive feature recognition model and reverse reconstruction with the help of sequential denoise model. At the same time, we eliminate the user-sensitive feature bias embedded in multi-interests to assist recommendation, and ultimately improve the fairness and personalization of the model through two-stage training. 
Extensive experiments on three real-world datasets demonstrate the effectiveness of FairGENRec on promoting recommendation accuracy and fairness.

\bibliographystyle{unsrt}  
\bibliography{references}

\end{document}